\documentclass{article}

\usepackage{hyperref}
\usepackage{url} 
\usepackage{xurl}

\usepackage{graphicx} 
\usepackage{amsmath}
\usepackage{subcaption}
\usepackage{caption}
\usepackage[export]{adjustbox}
\usepackage{comment} 
\usepackage{tikz}

\title{The Inverse Velocity Force and Automotive Physics}
\author{Chris L. Lin }
\date{\footnotesize Department of Physics, University of Houston, Houston, TX 77204-5005 \\ \today}

\begin{document}

\maketitle

\section*{}
\noindent
\begin{center}\LARGE
\textbf{Abstract}
\end{center}
 \vspace{\baselineskip}
The inverse velocity force $F(v)=C/v$ is central to the analysis of automotive performance, but is often unmentioned in classical mechanics courses. It represents the maximum force that can be delivered from a power-limited engine over a wide range of speeds. We show how such a force arises from a device called a transmission in internal combustion engines, and more simply from circuit controls in electric vehicles. We also examine how the force needs to be modified and supplemented to describe vehicle acceleration throughout the entire range of speeds, along with the ensuing kinematics. Along the way we compare gasoline vehicles with electric ones and derive the shape of their torque-speed curves.


\begin{center} \LARGE
\textbf{The Inverse Velocity Force and Automotive Physics}
\end{center}

\section{Introduction}
Over the course of their studies, students of physics solve problems for a variety of forces, not only developing a strong physical intuition for each force but also gaining an understanding of how each force is realized in the physical world. With this in mind we believe not enough emphasis is placed on the inverse velocity force $F(v)=C/v$ that is central to vehicular motion. Such a force delivers constant power, $P(t)=F(v(t))v(t)=C$, and represents the maximum force that can be extracted at any speed from a power-limited device. In this paper we explain how such a force arises from using a transmission in gasoline-powered vehicles (Sec. \ref{graphicalInterpretation}) and circuit controls in electric vehicles  (Sec. \ref{electricVehicles}) and explore the consequences of this force on automotive performance (Secs. \ref{ConstantPower}--\ref{Automotive}).


\section{Transmissions} \label{graphicalInterpretation}

Suppose we are given a device that can produce the force $F(v)$ at speed $v$, but are unsatisfied with this force function and would like a different one, but without extensively modifying the device or supplementing it with an additional source of energy to do so. If we connect it to the input port of a simple machine of mechanical advantage $N$, the simple machine will produce a force $\tilde{F}=N F(v)$ at speed $\tilde{v}=v/N$ at its output port. Expressing the output force $\tilde{F}$ as a function of the output speed $\tilde{v}$ we obtain

\begin{equation} \label{transformeqn}
    \tilde{F}(\tilde{v})=N F (N\tilde{v}).
\end{equation}

\begin{figure}
\renewcommand{\thesubfigure}{\alph{subfigure}}
     \begin{subfigure}{0.7\textwidth}
     \phantomcaption\label{constEngineForce}(\thesubfigure)\hfill
         \includegraphics[width=\linewidth,valign=t]{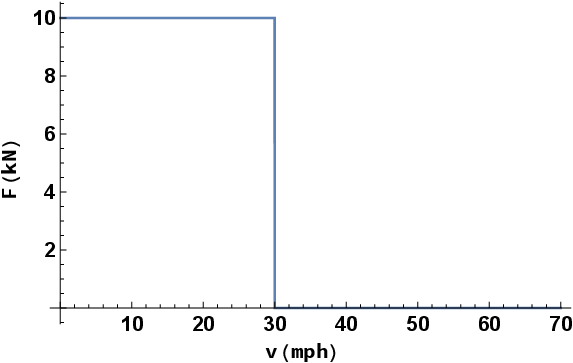}
     \end{subfigure}
     
     \begin{subfigure}{0.7\textwidth}
        \phantomcaption\label{constTransmissionForce}(\thesubfigure)\hfill
         \includegraphics[width=\linewidth,valign=t]{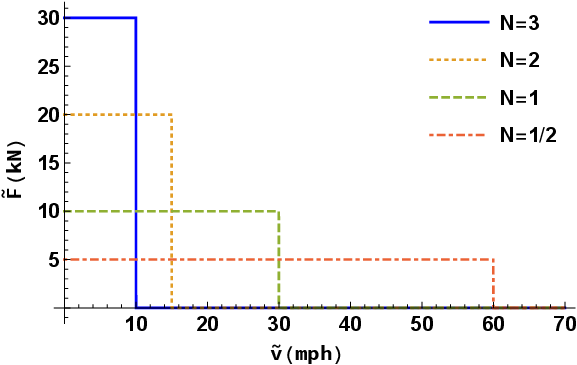}
     \end{subfigure}
     
\begin{subfigure}{0.7\textwidth}
        \phantomcaption\label{constComboForce}(\thesubfigure)\hfill
         \includegraphics[width=\linewidth,valign=t]{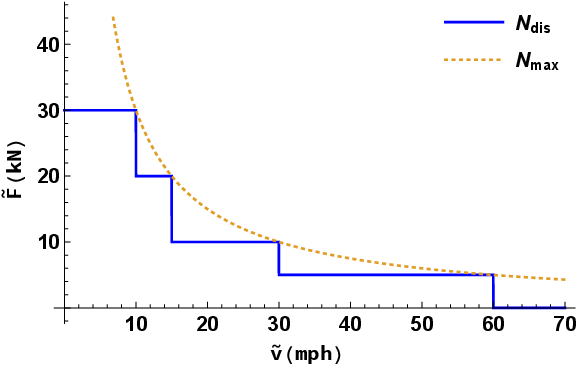}
     \end{subfigure}
     \caption{The input force function $F(v)$ in (a) is transformed into the output force function $\tilde{F}(\tilde{v})$ in (b) by use of simple machines, shown for different values of $N$. By varying the mechanical advantage $N=N(\tilde{v})$, a variety of transformations can be attained (c).}
          \label{constForceModel}
\end{figure}

Fig. \ref{constEngineForce} gives an example of an input force function that has a limited range punctuated by a sharp velocity cutoff. The output force function for a variety of values of $N$ is given in Fig. \ref{constTransmissionForce}. $N>1$ increases the force while decreasing its speed range while $N<1$ decreases the force while increasing its speed range.\footnote{This must be true as the transformation preserves the area: $\int_0^\infty Nf(Nx)dx =\int_0^\infty f(x) dx$.} \\

A \textit{transmission} is a device that allows $N$ to have different values for different speeds, i.e., $N=N(\tilde{v})$. By having such freedom a variety of transformations are possible. For example, the discrete selection

\begin{align}
N_{\text{dis}}(\tilde{v})=
\begin{cases}
    3, &\tilde{v}<10\\
    2, &10\leq \tilde{v}<15\\
    1, &15 \leq \tilde{v}<30\\
    1/2, &\tilde{v}\geq 30
\end{cases} \label{eqn2}
\end{align}

creates the staircase-shaped force function shown in Fig. \ref{constComboForce}, when $F(v)$ is as shown in Fig. \ref{constEngineForce}. When $N(\tilde{v})$ is piecewise constant like this we call each different piece a selected \textit{gear}. \\

Suppose the original force function $F(v)$ has maximum power delivery at $v^*$, i.e. $P(v)=F(v)v$ is maximized at $v=v^*$. Then there is a universal choice for $N(\tilde{v})$ (continuous rather than piecewise) that not only produces the $C/\tilde{v}$ force but delivers the maximum possible force achievable by any transmission: 

\begin{equation}
    N_{\text{max}}(\tilde{v})=\frac{v^*}{\tilde{v}},
\end{equation}

which, plugging into Eq. \eqref{transformeqn} causes the transformed force to become

\begin{equation}\label{maxForceExpress}
        \tilde{F}_{\text{max}}(\tilde{v})=\frac{v^*F (v^*) }{\tilde{v}} =\frac{P_{\text{max}}}{\tilde{v}}.
\end{equation}

Since 

\begin{equation} \label{fundamentalTransmissEqn}
   \tilde{F}(\tilde{v})=\frac{N \tilde{v}F (N\tilde{v})}{\tilde{v}}\leq \frac{P_{\text{max}}}{\tilde{v}},
\end{equation}

Eq. \eqref{maxForceExpress} represents the maximum force at each velocity that can be delivered from a power-limited engine, attained using such an ideal transmission $ N_{\text{max}}(\tilde{v})$. Such a force is shown by the dotted hyperbola in Fig. \ref{constComboForce}. Note that the force in Fig. \ref{constComboForce} is greater than or equal to the original force in Fig. \ref{constEngineForce} at every speed, and is nonzero over a wider range of speeds.\\

\begin{figure}
    \centering
    \includegraphics[width=0.8\linewidth]{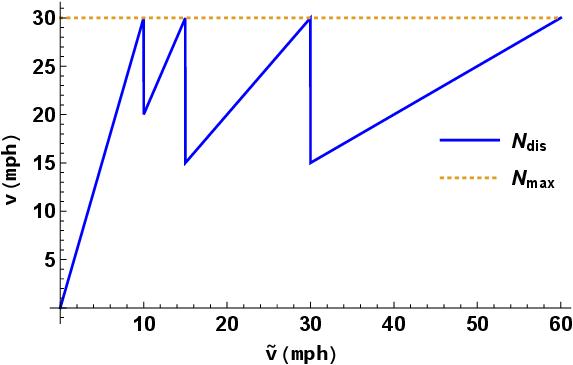}
    \caption{A plot of input velocity $v=N(\tilde{v})\tilde{v}$ against output velocity $\tilde{v}$ for two different transmissions. Sudden gear changes produce discontinuities in $v$.}
    \label{inputvelocvsoutputveloc}
\end{figure}

Since $v=N_{\text{max}}(\tilde{v})\tilde{v}=v^*$, the ideal transmission is always selecting the mechanical advantage that allows the input to be driven at $v^*$ no matter the value of $\tilde{v}$, so that maximum power is always delivered. Fig. \ref{inputvelocvsoutputveloc} shows how the engine speed $v$ varies with output speed $\tilde{v}$ for the discrete transmission defined by Eq. \eqref{eqn2}.\footnote{In the context of automobiles, Fig. \ref{inputvelocvsoutputveloc} would be analogous to a plot of the tachometer vs speedometer readings, with the tachometer dropping with each gear change to a lower $N$.}\\

Using these graphs we are ready to show how an inverse velocity force comes about: the discontinuous force function in Fig. \ref{constComboForce} can be approximated by the highest mechanical advantage gear followed by using the maximum force hyperbola of Eq. \eqref{maxForceExpress} to approximate the rest of the gears, resulting in a force function similar in shape to the one shown in Fig. \ref{effectivemodele}. The singularity at $\tilde{v}=0$ is resolved by the existence of a maximum mechanical advantage.

\section{Constant Power Kinematics} \label{ConstantPower}

The purpose of maximizing the force is to obtain the fastest acceleration, which was found to occur at constant power. In this section we examine the constant power equations of motion. Integrating $P=\frac{d}{dt}\left(\frac{1}{2}m v^2 \right)$ one gets 

\begin{equation} \label{constPowEqn2}
    v=\sqrt{v_i^2+\frac{2P}{m}(t-t_i)},
\end{equation}

while differentiating and integrating this expression gives
\begin{align}\label{constPowEqn3}
    a&=\frac{P}{m}\sqrt{\frac{1}{v_i^2+\frac{2P}{m}(t-t_i)}} \\ \nonumber
    x&=\frac{m}{3P}\left(v_i^2+\frac{2P}{m}(t-t_i)\right)^\frac{3}{2}+\left(x_i-\frac{m}{3P}v_i^3\right).
\end{align}

 The kinematics of the $P/v$ force can be better understood by comparing it to the more familiar constant force at very large times. The result is summarized in Tbl. \ref{forceComparison}. We note that although the acceleration falls to zero with time, it does not fall fast enough for the velocity to asymptote. Indeed, the stream of energy delivered by the force is constant, so does not even decrease with time.\\

\begin{table}[h]
\renewcommand{\arraystretch}{1.9} 

\begin{center}
\begin{tabular}{ |c| c |c| } 
\hline
  & $F=F_0$ & $F=\frac{P}{v}$ \\ \hline
 $a(t)$ & $\frac{F_0}{m}$ & $\sqrt{\frac{P}{2m}}\, t^{-1/2}$ \\  \hline
 $v(t)$ & $\left(\frac{F_0}{m}\right) t$ & $\sqrt{\frac{2P}{m}} \, t^{1/2}$  \\  \hline
 $x(t)$ & $\frac{1}{2}\left(\frac{F_0}{m}\right) t^2$ & $\sqrt{\frac{8P}{9m}}\, t^{3/2}$ \\ \hline
 $P(t)$ & $m\left(\frac{F_0}{m}\right)^2 t$ & $P$ \\ \hline
 
\end{tabular}\caption{Large $t$ behavior comparison between the constant force and constant power equations of motion. Besides the differing $t$ dependence, the determining kinematic scale switches from $F_0/m$ to $P/m$ \cite{10.1119/1.12897}.}
\label{forceComparison}
\end{center}
\end{table}

However, if we include dissipation, a maximum velocity will be reached. For automobiles, there exist approximate formulas for dissipation that include both rolling friction and air resistance \cite{10.1119/1.10552}. Formulas for specific vehicles come from the manufacturers themselves, who by law must submit to the Environmental Protection Agency (EPA) the dissipative force that each of their vehicles experience in the form $F_d(v)=A+Bv+Cv^2$ as part of their fuel economy calculations.\footnote{The database by model year can be accessed at \url{https://www.epa.gov/compliance-and-fuel-economy-data/data-cars-used-testing-fuel-economy}}  The coefficients $A$, $B$, $C$, known as the target coefficients, are experimentally determined using a coastdown test similar to the procedure outlined in \cite{10.1119/1.2341300}. The maximum velocity is then determined by a balance between $P/v$ and $F_d$:
 
 \begin{equation} \label{dissipation}
 \frac{P}{v_{\text{max}}}-A-Bv_{\text{max}}-Cv_{\text{max}}^2=0.
 \end{equation}

 For example, the EPA database lists the 2025 Aston Martin DB12 as having $A=53.24 \, lb \,(236.82\, N)$, $B=0.2228 \, lb/mph \,(0.6158\, N/kph)$, and $C=0.0221 \,lb/mph^2 \,(0.0380\, N/kph^2) $. With an advertised maximum power of $P=671 \,hp\, (500\, kW)$, $v_{\text{max}}$ is then calculated to be $218 \, mph \,(351\, kph)$, which is higher than the advertised top speed of $202 \, mph \,(325\, kph)$ \cite{carndriver}. Conversely, given $v_{\text{max}}=202 \, mph \,(325\, kph)$, one can from Eq. \eqref{dissipation} calculate a realized power of $539 \, hp \,(402\, kW)$. The discrepancy between the advertised power and the calculated power arises because the manufacturer measures power directly at the engine rather than at the wheels, which neglects frictional and inertial losses along the drivetrain\footnote{Inertial losses only become significant while in 1st gear due to its high gear ratio \cite{buschert2003measuring}.} and can be as high as $20 \%$ of the engine power \cite{inproceedings}.

\section{Application to Automobiles} \label{Automotive}

For automobiles the arguments in Sec. \ref{graphicalInterpretation} proceed as before with the torque $\tau$ replacing the force $F$ and the angular velocity $\omega$ replacing the velocity $v$. In particular, $\tau(\omega)$ is an input torque from an engine crankshaft rotating at speed $\omega$ and $\tilde{\tau}(\tilde{\omega})$ is the transmission output torque applied to wheels spinning at speed $\tilde{\omega}$. One gets the angular counterpart to Eq. \eqref{fundamentalTransmissEqn}:

\begin{align} \label{angularfundamentalTransmissEqn}
   \tilde{\tau}(\tilde{\omega})= N(\tilde{\omega})\tau(N(\tilde{\omega})\tilde{\omega})\leq \frac{P_{\text{max}}}{\tilde{\omega}}.
\end{align}

\begin{figure}
\centering 
\begin{tikzpicture}
\draw[thick]  (0,1) circle (1);
\draw[thick]  (0,1)--(1,1) node[midway, sloped, above] {$R$};
\draw[->,thick] (0,0)--(-1,0) node[midway, sloped, below] {$f'_j$};
\filldraw[black] (0,0) circle (2 pt); 
\draw[thick]  (3,1) circle (1) node{$\tilde{\tau}_i$};
\draw[->,thick] (3,0)--(4,0) node[midway, sloped, below] {$f_i$};
\draw[->,thick] (2.5, 1) arc(180:-160:.5);
\filldraw[black] (3,0) circle (2 pt); 
\end{tikzpicture}
        \caption{Free-body diagrams for the undriven and driven wheels of a car accelerating to the right. The transmission output torque and the torque from friction accelerate the driven and undriven wheels in the clockwise direction, respectively, while friction on the driven wheels accelerates the car to the right.} 
        \label{picturetorqueToForce} 
\end{figure}

To proceed we need an expression for how the torque at the wheels translates to a force on the car. For a car of mass $M$ rolling without slipping with acceleration $a=R \alpha$, the equations of motion resulting from the free-body diagrams in Fig. \ref{picturetorqueToForce} are

\begin{align}\label{wheelEqn}
 \sum_i f_i- \sum_j f'_j&=M (R \alpha)\\ \nonumber
    \tilde{\tau}_i-Rf_i&=I_i \alpha\\ \nonumber
    Rf'_j&=I'_j \alpha,
\end{align}    

\noindent where $R$, $I$, and $\alpha$ are the radius, moment of inertia, and angular acceleration of the wheels, respectively, $\tilde{\tau}$ the wheel torque and $f$ the static frictional force, and the primed variables and index $j$ are used to distinguish the undriven from driven wheels. Multiplying the top line of Eq. \eqref{wheelEqn} by $R$ followed by adding all the lines together solves for $\alpha$, and insertion of $\alpha$ back to the top line gives the net force on the car to be:

\begin{align}\label{torqueToForce}
    \tilde{F}(\tilde{v})=\frac{\tilde{\tau}(\tilde{\omega})}{R\left(1+I_{\text{wheels}}/M R^2\right)}\approx\frac{\tilde{\tau}(\tilde{\omega})}{R},
   \end{align}

where $\tilde{\tau}$ and $I_{\text{wheels}}$ are the sum of the torques and moments of inertia over all the wheels, respectively, and $\tilde{v}=R \tilde{\omega}$ \cite{10.1119/1.5051145}. Substituting Eq. \eqref{angularfundamentalTransmissEqn} into Eq. \eqref{torqueToForce} gives

\begin{align}\label{torqueGeareqn}
    \tilde{F}(\tilde{v})=\frac{1}{R}N(\tilde{\omega})\tau(N(\tilde{\omega})\tilde{\omega})\leq \frac{P_{\text{max}}}{\tilde{v}}.
   \end{align}

In order to apply Eq. \eqref{torqueGeareqn} exactly, we will need the torque-speed curve for an electric motor or an internal combustion engine, along with their gearing. We do this in Sec. \ref{electricVehicles} for an electric car and in Appx \ref{appendixCombustion} for a car with a gasoline engine. For the rest of this section, we will instead approximate the LHS of Eq. \eqref{torqueGeareqn} with its upper bound so that only the single number $P_{\text{max}}$ needs to be specified. \\

We have seen in Sec. \ref{graphicalInterpretation} that approximating $ \tilde{F}(\tilde{v})$ by $P_{\text{max}}/\tilde{v}$ can fail at low $\tilde{v}$ due to the existence of a maximum possible mechanical advantage. There is, however, another factor that invalidates this approximation for low $\tilde{v}$. We note that in the derivation of Eq. \eqref{torqueToForce}, we have assumed that static friction is large enough to prevent slipping. However, static friction is constrained by $f_i \leq \mu_s N_i$, where $\mu_s$ is the coefficient of static friction and $N_i$ is the normal force on the $i$th driven wheel. Plugging this restriction into Eq. \eqref{wheelEqn} gives a maximum force of

\begin{align}\label{maxFrictionForce1}
    \tilde{F}(\tilde{v}) \leq \frac{\mu_s\sum_i N_i}{1+(\sum_j I'_j)/MR^2} \approx \mu_s\sum_i N_i \equiv \tilde{F}_{\text{max}}.
\end{align}

For example, a two-wheel drive car with a uniform weight distribution would have a maximum force of $\mu_s M g/2$, while a four-wheel drive car would have a maximum force of $\mu_s Mg$. To be more accurate, we would need to take into account the actual weight distribution of a car along with shifts in $N_i$ that occur when accelerating: formulas for $N_i$ incorporating such effects can be found in \cite{10.1119/1.16430}.\\

\begin{figure}
\centering 
\includegraphics[width=0.5\linewidth]{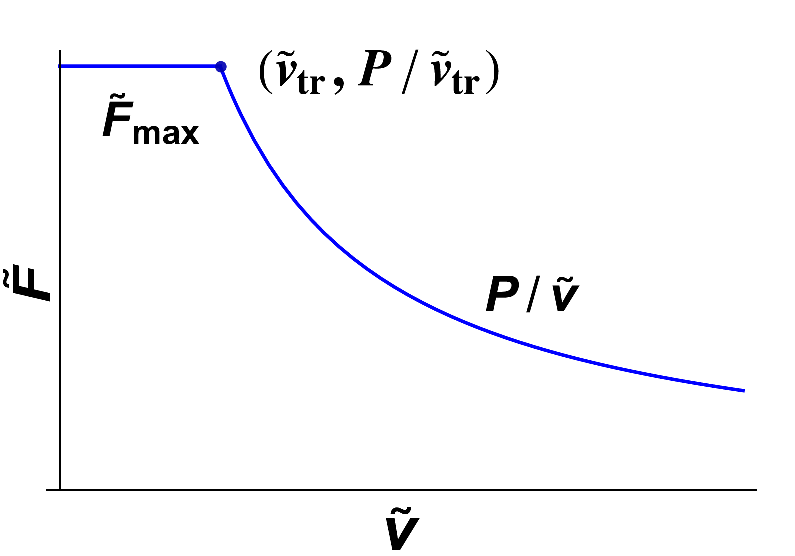}
    \caption{A model for propulsive force, exhibiting a transition from a traction-limited, constant force regime to an engine-limited, constant power regime.}
    \label{effectivemodele}
      \end{figure}

We now have a force model that can be taken as the starting point for ``0-60'' analysis \cite{10.1119/1.16430,10.1119/1.1987216,10.1119/1.16893} in car racing, as shown in Fig. \ref{effectivemodele}. The transition point from a traction-limited, constant force regime to an engine-limited, constant power regime occurs at

\begin{equation}
\tilde{v}_{\text{tr}}= \frac{P}{\tilde{F}_{\text{max}}}.
\end{equation}

It is straightforward to determine all the remaining values at the transition point. For constant acceleration $a= \tilde{F}_{max}/M$ the velocity $\tilde{v}_{tr}$ is reached at time

 \begin{equation}
t_{\text{tr}}=\frac{M P}{\tilde{F}_{\text{max}}^2},
 \end{equation}
 
after which the car will have traveled a distance

\begin{equation}
    x_{\text{tr}}=\frac{M P^2}{2 \tilde{F}_{\text{max}}^3 }.
\end{equation}

These values can be inserted into Eq.'s \eqref{constPowEqn2} and \eqref{constPowEqn3} as the initial conditions $(x_i ,v_i, t_i)=(x_{\text{tr}} ,\tilde{v}_{\text{tr}}, t_{\text{tr}})$  for analysis in the constant power regime.\\

\section{Electric Vehicles}\label{electricVehicles}

\begin{figure}
\renewcommand{\thesubfigure}{\alph{subfigure}}
     \begin{subfigure}{0.7\textwidth}
      \phantomcaption\label{evtorquespeed}(\thesubfigure)\hfill
         \includegraphics[width=\linewidth,valign=t]{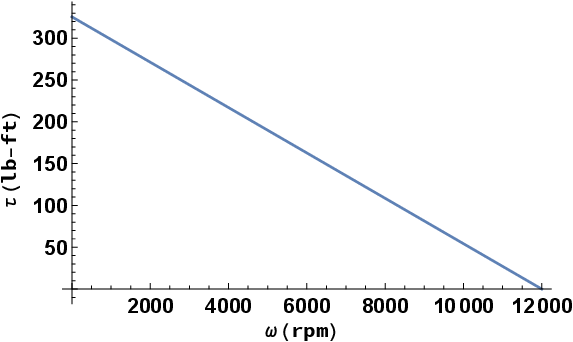}
     \end{subfigure}
     \begin{subfigure}{0.7\textwidth}
       \phantomcaption\label{plotcomparison1}(\thesubfigure)\hfill
         \includegraphics[width=\linewidth,valign=t]{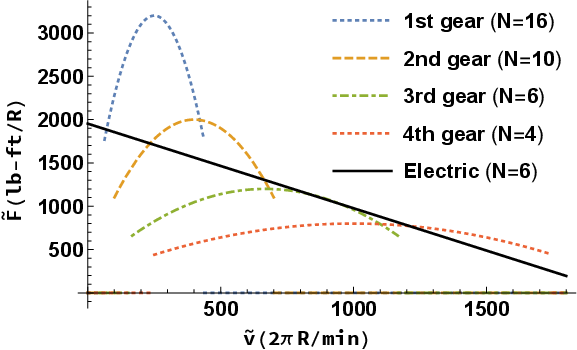}
     \end{subfigure}
     \begin{subfigure}{0.7\textwidth}
         \phantomcaption\label{plotcomparison2}(\thesubfigure)\hfill
         \includegraphics[width=\linewidth,valign=t]{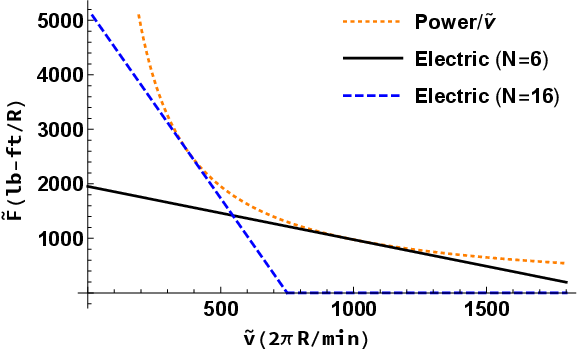}
     \end{subfigure}
     \caption{The torque-speed curve of the electric motor shown in (a) results in the propulsive force shown in (b) with the use of a single $N=6$ gear. For comparison, force curves for a gasoline vehicle for several values of $N$ are overlaid. A few electric vehicles offer a two-speed transmission as shown in (c).}
     \label{electricVehiclegraphs}
\end{figure}

It is instructive to consider the utility of transmissions in electric vehicles, a rapidly emerging\footnote{Electric vehicles actually predate internal combustion engines \cite{secondcarndriver}, but only recently has the technology and cost improved enough to present an economically-viable alternative.} technology that currently comprises one-fifth of global car sales \cite{IEA}. Fig. \ref{evtorquespeed} gives the torque-speed curve of an electric vehicle running on a DC motor with a maximum power of $186 \,hp$ and a maximum speed of $12000\, rpm$, modeled using Eq. \eqref{newreparameterization} in Appx. \ref{appendixElectric}. The large value for the maximum speed, almost double that seen in a standard gasoline engine, is typical of the electric motors used in automotive applications. Applying Eq. \eqref{torqueGeareqn} to this curve for $N=6$ results in Fig. \ref{plotcomparison1}, where for comparison we have superimposed the curves shown in Fig. \ref{phaseTransitionModel1} for the gasoline engine described in Appx \ref{appendixCombustion}. One can see that due to the wide rpm range of the motor and the low value of $N=6$, only a single gear is needed to span the entire range of speeds, while four gears are required for a combustion engine to navigate over the same range. Indeed, most electric cars do not have a multi-speed transmission and only require this single fixed gear. A few electric cars offer a two-speed transmission: this is illustrated in Fig. \ref{plotcomparison2} where the gear change is made at the intersection of the $N=16$ and $N=6$ curves. The $N=16$ gear provides a lot of torque but has a limited speed range, while the $N=6$ gear provides a wide speed range. \\

\begin{figure}
\renewcommand{\thesubfigure}{\alph{subfigure}}
     \begin{subfigure}{0.7\textwidth}
      \phantomcaption\label{induction}(\thesubfigure)\hfill
         \includegraphics[width=\linewidth,valign=t]{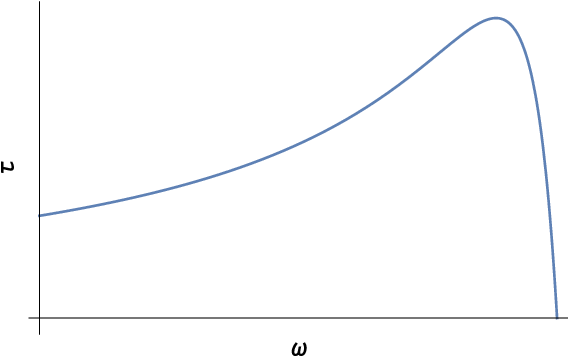}
     \end{subfigure}
     \begin{subfigure}{0.7\textwidth}
       \phantomcaption\label{controlinduction}(\thesubfigure)\hfill
         \includegraphics[width=\linewidth,valign=t]{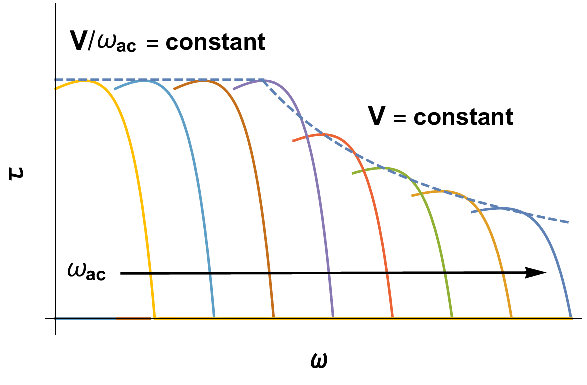}
     \end{subfigure}
     \caption{The torque-speed curve for an induction motor operating at a fixed voltage $V$ and frequency $\omega_{\text{ac}}$ is shown in (a). Varying these two parameters with motor speed $\omega$ results in the dashed torque-speed curve shown in (b). Increasing $\omega_{\text{ac}}$ displaces the curve in (a) to the right, where in the constant torque region $V$ is also increased so that the ratio $V/\omega_{\text{ac}}$ remains constant, and in the constant power region $V$ attains its maximum value and only $\omega_{\text{ac}}$ increases.}
     \label{inductionMotorgraphs}
\end{figure}

Finally, we note that the three-phase AC induction motor, the most widely used motor for industrial applications, has a torque-speed curve that by itself is not suitable for automotive applications, as shown in Fig. \eqref{induction} \cite{Shayak_2019}. However, this torque-speed curve only applies for fixed values for the voltage $V$ and AC frequency $\omega_{ac}$ applied to the motor.\footnote{ We note that for induction motors, the speed of the motor $\omega$ need not equal the frequency $\omega_{\text{ac}}$ driving the motor. The difference between these two quantities is measured by a parameter known as ``slip.''} Using a variable frequency and voltage drive, along with a sensor to measure the motor speed $\omega$, both voltage and frequency can be adjusted based on motor speed so that $V=V(\omega)$, $\omega_{\text{ac}}=\omega_{\text{ac}}(\omega)$, and  $\tau(\omega,V(\omega),\omega_{\text{ac}}(\omega))=\tau_{\text{ideal}}(\omega)$, where $\tau_{\text{ideal}}(\omega)$ is the ideal torque-speed curve for electric vehicles as shown by the dashed line in Fig. \ref{controlinduction}. The functions $\omega_{\text{ac}}(\omega)$  and $V(\omega)$ needed for this conversion can be found in \cite{gao}. Therefore increasing $\omega_{\text{ac}}$ as the car accelerates is the electrical equivalent of decreasing $N$ for an internal combustion engine as it accelerates, as both operations shift the original torque-speed curve in order to expand the range for use in automotive applications. Given the flexibility of electrical circuits over mechanical parts in designing control systems, it is not surprising that a variety of electric motors can be controlled to have their torque-speed curve in the shape of $\tau_{\text{ideal}}(\omega)$, obviating the need for a mechanical transmission. A few of the techniques for doing this are discussed in Appx. \ref{appendixElectric}.

\section{Conclusion} \label{conclusion}

The inverse velocity force $F(v)=C/v$ is not often included in introductory physics courses. Besides being a rare example of a velocity-dependent force that is not frictional, it has the unique property of transmitting constant power. Abstractly it represents the greatest force you can extract from a power-limited device, and practically it finds application in automotive performance. For gas-powered vehicles, a multi-gear mechanical transmission is used to approximate an inverse velocity force, while in electric vehicles, it arises more simply from circuit controls. We hope to have shown that the inverse velocity force is worth adding to the list of forces that physics students have in their toolkit.  

\section*{Acknowledgments}

The author would like to acknowledge the reviewers for their helpful suggestions.

\newpage

\appendix

\begin{center}
\huge{\textbf{Appendices}}
\end{center}

\section{Torque-Speed Curve for Electric Motors} \label{appendixElectric}

\begin{figure}
\renewcommand{\thesubfigure}{\alph{subfigure}}
     \begin{subfigure}{0.7\textwidth}
       \phantomcaption\label{circuitMotor}(\thesubfigure)\hfill
         \includegraphics[width=\linewidth,valign=t]{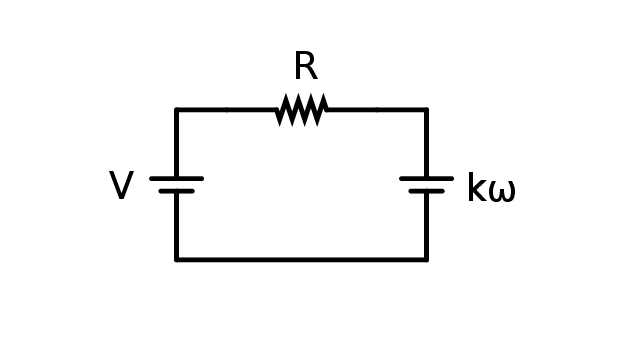}
     \end{subfigure}
     \begin{subfigure}{0.7\textwidth}
      \phantomcaption\label{torqueCurveELectric}(\thesubfigure)\hfill
         \includegraphics[width=\linewidth,valign=t]{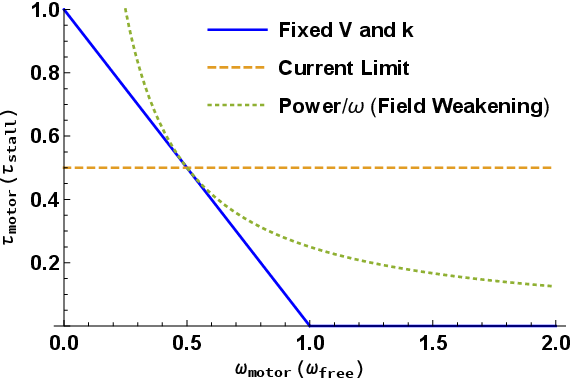}
     \end{subfigure}
     \caption{The circuit diagram for the DC motor shown in (a) results in the torque-speed curves shown in (b). Allowing the voltage $V$ and the motor constant $k$ to vary with motor speed $\omega$ gives the dashed and dotted curves.}
     \label{ElectricCarCharacteristics}
\end{figure}

The simplest motor that captures the torque-speed curve of an electric vehicle is the permanent magnet brushed DC motor, consisting of a loop of wire placed in the magnetic field created by the permanent magnet. When a current $I(t)$ flows through this loop, a torque $\tau(t)=\eta I(t)$ is exerted on it, where the torque constant $\eta$ depends on the geometry of the loop and the strength of the magnetic field. As the loop rotates in the magnetic field, it encounters a back-emf proportional to its speed, $\Delta V_{\text{motor}}=-k \omega$ as shown in Fig. \ref{circuitMotor}, where $R$ is the resistance of the loop and $k$ is numerically equal to $\eta$.\footnote{A simple argument for the equality of $k$ and $\eta$ is that $I\,\Delta V_{\text{motor}}=(\tau/\eta)(-k \omega)$, so for there to be complete conversion of electrical to mechanical power, $k=\eta$.}  \\

From Kirchoff's rules we have $V-IR-k \omega=0$, and substituting $I=\tau/k$ into this expression gives:

\begin{align}\label{reparameterization}
\tau&=\frac{Vk}{R}-\left(\frac{k^2}{R}\right)\omega \\ \nonumber
&=\tau_{\text{stall}}-\left(\frac{\tau_{\text{stall}}}{\omega_{\text{free}}}\right)\omega,
\end{align}

where $\tau_{\text{stall}}=Vk/R$ is the torque at zero speed and $\omega_{\text{free}}=V/k$ is the speed at zero torque. We have rewritten $\tau(\omega)$ in terms of mechanical parameters rather than electrical ones, the result of which is the solid line in Fig. \ref{torqueCurveELectric}. \\

The maximum power for Eq. \eqref{reparameterization} is $P_{\text{max}}=\tau_{\text{stall}}\,\omega_{\text{free}}/4=V^2/4R$ and occurs at $\omega=\omega_{\text{free}}/2$. Using this we can further rewrite Eq. \eqref{reparameterization} as

\begin{align}\label{newreparameterization}
\tau(\omega)=\frac{4P_{\text{max}}}{\omega_{\text{free}}} -\left(\frac{4 P_{\text{max}}}{\omega_{\text{free}}^2}\right)\omega.
\end{align}

This simple model predicts that the torque-speed curve is a linearly decreasing curve. However, the motor draws the most current at low speeds when the back-emf is small, and to prevent overheating and motor burnout the current and therefore the torque needs to be restricted by adding a variable-voltage controller to the motor circuit, only ramping to the full voltage $V$ when the speed hence back-emf is large enough to prevent a large current. Such a restriction is shown by the dashed line in Fig. \ref{torqueCurveELectric}, whereby $\tau(\omega)$ follows the dashed line until it intersects the solid line, and follows the solid line thereafter.\footnote{High current can be tolerated for a short period, so if high torque is only briefly used (as in launching the car from rest), rather than a sustained hill climb, then the dashed horizontal line of Fig. \ref{torqueCurveELectric} can be made higher.}\\

If the permanent magnet is swapped for an electromagnet, the strength of the magnetic field (and therefore $k$) can be adjusted depending on the value of $\omega$ measured by a sensor. For large $\omega$, the magnetic field can be reduced by lowering the current in the electromagnet, which leads to a lower $k$ and a larger $\omega_{\text{free}}$. This technique is known as \textit{field weakening} and can extend the speed range of the motor after the maximum value of $V$ is attained at the end of the voltage ramp-up. Differentiating Eq. \eqref{reparameterization} w.r.t. $k$ and setting to zero one finds that $k(w)=V/2\omega$ will maximize the torque for a given $\omega$, where it attains the value $\tau(w)=(V^2/4R)(1/\omega)$ and recovers the inverse-velocity dependence. By weakening the field in this way, the motor delivers constant power beyond the current limit.\\

Although the use of DC motors in electric cars has been phased out, the torque-speed curves for AC motors are all similar to Fig. \ref{torqueCurveELectric} after controlling the voltage, magnetic field strength, and AC frequency as a function of motor speed. The shape of this curve is ideal in that a high constant torque is provided at lower speeds where it is needed, allowing cars to accelerate to cruising speed in a short period of time. Because every device is power limited, the torque must eventually drop at higher speeds, and the constant power curve has the minimum drop. This in part explains why mechanical transmissions do not benefit electric cars as much as they benefit internal combustion engines, as these benefits have already been achieved electronically.

\section{Internal Combustion Engines}\label{appendixCombustion}

\begin{figure}
     \renewcommand{\thesubfigure}{\alph{subfigure}}
     \begin{subfigure}{0.7\textwidth}
         \phantomcaption\label{phaseTransitionModel3}(\thesubfigure)\hfill
         \includegraphics[width=\linewidth,valign=t]{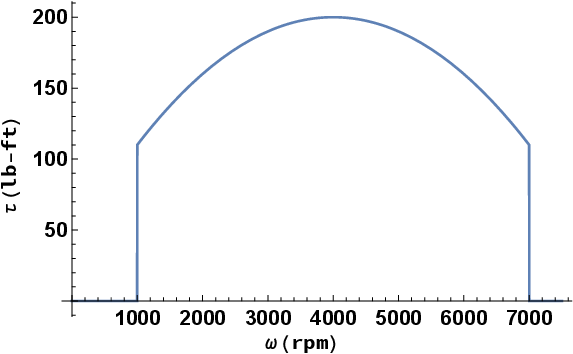}
     \end{subfigure}
     \begin{subfigure}{0.7\textwidth}
        \phantomcaption\label{phaseTransitionModel1}(\thesubfigure)\hfill
         \includegraphics[width=\linewidth,valign=t]{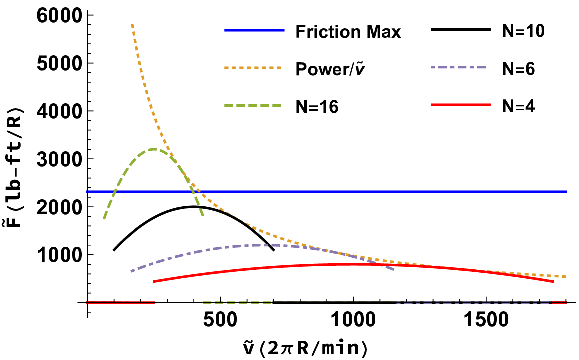}
     \end{subfigure}
     \begin{subfigure}{0.7\textwidth}
       \phantomcaption\label{phaseTransitionModel2}(\thesubfigure)\hfill
         \includegraphics[width=\linewidth,valign=t]{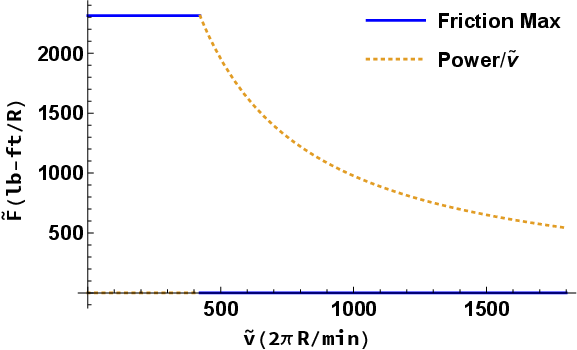}
     \end{subfigure}
     \caption{The torque of the combustion engine shown in (a) results in the propulsive force shown in (b) for different gear ratios, where we have also included the maximum force allowed by friction. Switching gears where the different force curves intersect approximates the inverse velocity force, resulting in the simplified model shown in (c).}
     \label{phaseTransitionModel}
\end{figure}

We begin by roughly describing how the overall shape of the torque-speed curve for an internal combustion comes about, using as an example the four-stroke spark-ignition engine. If we denote $E$ as the chemical energy released by each combustion event, and note that one combustion event occurs for every two revolutions of the crankshaft \cite{10.1119/10.0002372}, then the mechanical energy delivered in one complete cycle is $\eta E= \tau (4\pi)$, where $\eta$ is the efficiency of the thermodynamic Otto cycle \cite{Serway2018PhysicsFS,10.1119/1.2341182} and $\tau$ is the average torque over two revolutions of the crankshaft. The torque-speed curve is then given by $\tau=(1/4 \pi) \eta E$, suggesting the torque-speed curve is independent of $\omega$. However, a number of real-world effects can cause deviations from this flat profile, the most important of which have to do with the fuel intake process: the amount of energy released in each combustion event depends on the amount of air-fuel mixture drawn into the piston cylinder by the partial vacuum created during the induction stroke, which can depend on $\omega$, so that $E=E(\omega)$, giving rise to $\tau(\omega)=(1/4 \pi) \eta E(\omega)$. Some general factors influencing the shape of $E(\omega)$ can be found in \cite{Heywood2018}, but actual torque-speed curves need to be measured using a dynamometer or reconstructed from acceleration data \cite{buschert2003measuring,graney2005alternative}.\footnote{Measured torque-speed curves can be found online and in publications. For analysis, a computer program like PlotDigitizer can be used to convert the graph into data points, which can then be used numerically or fitted to a function with adjustable parameters.} For the purpose of illustrating Eq. \eqref{torqueGeareqn} however, we will consider a toy model with the simple function

\begin{align}
    \tau(\omega)=\left(200-\frac{40}{(2000)^2}\left(\omega-4000\right)^2\right)\theta(\omega-1000)\theta(7000-\omega),
\end{align}

where $\tau$ is measured in $lb \, \text{-}ft\,(1\, lb \, \text{-}ft=1.356\, N\text{-}m)$, $\omega$ in $rpm$, and $\theta$ is the Heaviside step function that imposes sharp cutoffs at $1000$ and $7000 \,\, rpm$: see Fig. \ref{phaseTransitionModel3}. This models a car with a peak torque of $200\, lb\, \text{-}ft$ at $4000 \, \,rpm$ that falls by $40 \,lb\, \text{-}ft$ when the rpm deviates by $2000\, \,rpm$ from its peak. The maximum horsepower of this torque function is $186 \,hp \,(1\,hp=746 \,W)$ and occurs at $5573\, \,rpm$. The cutoffs are used to enforce the limited stable operating range of internal combustion engines. \\

We take the gear ratios for the car to be $16$, $10$, $6$, and $4$,\footnote{The gearbox of the car will have the values $4$, $2.5$, $1.5$, and $1$, but a final drive value of $4$ applies to each gear selection, and the total gear ratio is the product of the two.} and the radius of the wheels to be $1\, ft$. For future reference we will also need to calculate the maximum static friction force. For a four-wheel drive, 1500 $kg$ car with coefficient of friction $0.7$, we get $F=\mu M g=10300\, N\, (2315 \,lb)$. \\

The plot of Eq. \eqref{torqueGeareqn} can now be constructed, as shown in Fig. \ref{phaseTransitionModel1}. If gear changes are made at the intersection of the gear curves, then a constant power hyperbola is approximated. However, there are a couple of subtleties at low speeds where $N=16$: first, there is a region where the calculated force exceeds the friction limit and second, a region before that where the calculated force abruptly falls to zero.\\

The latter problem indicates that the engine would be taken below its operating range, and because the engine speed and the wheel speed are coupled by the relation $\tilde{\omega}=\omega/N$, this occurs at wheel speed $1000/16 \,\, rpm$. To prevent this, instead of having the engine connect directly to the input of the transmission, a \textit{clutch} is inserted between them. The connection between the engine and the clutch is done through friction contact allowing them to rotate at different speeds by slipping. Above the cutoff wheel speed, there is no slipping and the torque and speed of the engine equals that of the clutch. Below the cutoff, the clutch can be disconnected from the engine, which can then be operated above its cutoff speed to produce a useable torque; engaging the clutch then brings the engine and transmission smoothly to the same speed.\footnote{To launch an automatic transmission, one holds the brake, revs the engine, then releases the brake. It should be noted that electric vehicles, with their ability to create torque from rest, are faster over shorter distances \cite{aaaaa} without having to wear out components.} We can replace all these complications with the simple model shown in Fig. \ref{phaseTransitionModel2}, where the force curve follows the maximum friction line of Fig. \ref{phaseTransitionModel1} until it intersects the $P/v$ curve, and traces it thereafter. Once the desired target velocity is reached and acceleration is no longer required, the transmission can then be used to select the gear that in combination with the gas pedal position allows the engine to operate where it conserves the most fuel rather than where it generates the most power, only needing enough power to overcome frictional forces.

\end{document}